# Electrocatalytic Hydrogen Peroxide Generation Using WO$_3$ Nanoparticle-Decorated Sodium Niobate Microcubes

Vanessa S. Antonin [1], Felipe M. Souza [2], Victor S. Pinheiro [1], João P. C. Moura [1],

Aline B. Trench [1], Caio Machado Fernandes [1],

Marcos R. V. Lanza [3], Mauro C. Santos [1*]

[1] *Centro de Ciências Naturais e Humanas (CCNH), Universidade Federal do ABC (UFABC). Rua Santa Adélia 166, Bairro Bangu, 09210-170, Santo André - SP, Brazil.0*

[2] *Departamento de química, Instituto Federal de Educação, Ciência e Tecnologia Goiano, BR-153, Km 633, Zona Rural, 75650-000, Morrinhos - GO, Brazil.*

[3] *Instituto de Química de São Carlos (IQSC), Universidade de São Paulo (USP), Avenida Trabalhador São-carlense 400, 13566-590, São Carlos, SP, Brazil.*

*Corresponding Author:*

*\*E-mail:* mauro.santos@ufabc.edu.br *(M. C. Santos).*

**Abstract**

The current work studies the electrocatalytic performance of $NaNbO_3$ microcubes decorated with $WO_3$ nanoparticles on Printex L6 at varying concentrations (1%, 3%, 5%, and 10% by weight) for $H_2O_2$ electrogeneration aiming for future use in electrochemical advanced oxidation processes for organic pollutant degradation. $H_2O_2$ electrogeneration was studied using oxygen reduction reaction (ORR) with the rotating ring-disk electrode (RRDE) technique. Electrochemical results revealed an improvement in $H_2O_2$ electrogeneration for the $NaNbO_3@WO_3/C$ materials compared to that achieved with Printex L6 carbon. Notably, the 5% $NaNbO_3@WO_3/C$ electrocatalyst exhibited a higher ring current for oxygen reduction reaction and promoted a 2.1-electron transfer, facilitating a higher rate of $H_2O_2$ electrogeneration through the 2-electron mechanism. Also, enhancing oxygen-containing functional groups has shown the capability to thoroughly adjust characteristics and enhance active sites, increasing $H_2O_2$ electrogeneration. These findings suggest that 5% $NaNbO_3@WO_3/C$ electrocatalysts hold promise for *in situ* hydrogen peroxide electrogeneration.

**1. Introduction**

Hydrogen peroxide (H$_2$O$_2$) is widely utilized as a cleaning oxidant across various sectors, including medical treatments, military applications, metallurgy, cosmetics, and chemical industries [1]; it has traditionally been produced through the anthraquinone (A.Q.) oxidation process in the industrial realm [2]. However, this method has long grappled with substantial energy consumption, stringent transportation limitations, and inherent shortcomings [3]. An appealing alternative to the industrial A.Q. approach is the oxygen reduction reaction (ORR), presenting an opportunity for decentralized H$_2$O$_2$ production. Within the ORR framework, H$_2$O$_2$ can be generated via a 2-electron reduction reaction via a cathodic electrolysis.

$$O_2 + 2H^+ + 2e^- \rightarrow H_2O_2 \qquad (1)$$

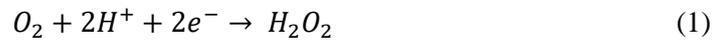

The ORR occurs through gas diffusion in the reaction medium, adsorbing on the active surface of electrocatalysts. Subsequently, electron transfer takes place, followed by product desorption. Finally, the release of active sites, with the electrocatalyst's catalytic activity, depends on its composition [4]. Carbonaceous materials are well-established in the literature for H$_2$O$_2$ electrogeneration via the ORR with a $2e^-$ pathway, such as graphite, carbon nanotubes, activated carbon fibers, carbon filters, and gas-diffusion electrodes (with O$_2$ or atmospheric air) made of carbon-polytetrafluoroethylene (P.T.F.E.) [4]. To maximize H$_2$O$_2$ electrogeneration by gas-diffusion electrodes (GDEs) using carbon as a catalyst, some nanostructured materials have been embodied into the carbon, enhancing their catalytic activity [5–10].

The perovskite structure of sodium niobate (NaNbO$_3$) exhibits a cubic arrangement and possesses intriguing characteristics, including straightforward synthesis, affordability, elevated crystallinity, reduced density, favorable chemical stability, and minimal environmental impact [11]. Given its distinctive configuration and photocatalytic attributes, NaNbO$_3$ has been found to be applicable in the reduction of CO$_2$ and the degradation of dyes [11]. It is important to note that Brazil has the largest niobium reserves, holding approximately 98% of the world's niobium [12], underscoring the need to expand niobium applications in the country. Using NaNbO$_3$ electrocatalysts combined with semiconductors or noble metals has shown promise in enhancing the electrocatalytic activity in some applications [12–15]. Antonin *et al.* [16] investigated the catalytic efficacy of NaNbO$_3$@CeO$_2$/C materials in the electrogeneration of H$_2$O$_2$. These materials displayed significant potential in the context of the ORR, exhibiting a preference for the 2-electron mechanism and resulting in an increased rate of H$_2$O$_2$ electrogeneration. Density functional theory calculations revealed that NaNbO$_3$ and CeO$_2$ surfaces possess a similarly low theoretical overpotential for this reaction, and CeO$_2$ enhances the catalyst by promoting electron transfer [16].

Tungsten trioxide (WO$_3$) acts as an n-type semiconductor. It has notably high electrical conductivity [17], holding scientific and technological significance due to its optical and electronic characteristics, which have a bearing on photocatalysis, sensors, and electrochromic devices[18]. The catalytic activity of tungsten oxide in the ORR has already been studied. Assumpção et al. [19] investigated the performance of WO$_3$ as an electrocatalyst for H$_2$O$_2$ electrogeneration via ORR They confirmed that mixing a small amount of carbon significantly enhanced Vulcan carbon's catalytic efficiency. This metal oxide presents surface acidity owing to Lewis-Brønsted acidic sites associated with W$^{6+}$ species, resulting in increased hydrophilicity and superior ORR results for H$_2$O$_2$ electrogeneration [19].

Thus, NaNbO$_3$@WO$_3$ hetero-structures have been effectively produced using a straightforward synthesis approach. The analysis of the materials included examinations of their morphology, crystalline phases, and surface composition through techniques such as scanning electron microscopy (S.E.M.), X-ray powder diffraction (XRD), and X-ray photoelectron spectroscopy (XPS). At the same time, the wettability of each catalyst was assessed through contact angle measurements. The study explores the potential of NaNbO$_3$@WO$_3$-based electrocatalysts with varying carbon proportions (w/w) to enhance their electrocatalytic properties towards H$_2$O$_2$ generation. This electrogeneration was assessed via a rotating ring-disk electrode (RRDE). The findings demonstrate that NaNbO$_3$@WO$_3$/C composites exhibit superior electrocatalytic activity compared to 4% NaNbO$_3$/C and pure carbon. The enhanced performance can be ascribed to increased oxygen acid groups on the Printex L6 surface, resulting from the nanostructured material NaNbO$_3$@WO$_3$/C. This augmentation favors the overall surface's hydrophilicity and increases H$_2$O$_2$ electrogeneration.

## 2. Experimental

### 2.1. Chemicals

All reagents, including sodium hydroxide (NaOH, Synth), hydrochloric acid (HCl, Synth), niobium oxide (Nb$_2$O$_5$, Sigma-Aldrich), tungsten chloride IV (WCl$_4$, 95%, Sigma-Aldrich), polyvinylpyrrolidone (PVP, 99%, Sigma-Aldrich), ascorbic acid (99%, Sigma-Aldrich), carbon black Printex L6 (PL6, Evonik Brazil), sodium sulfate (Na$_2$SO$_4$, Sigma-Aldrich) sulfuric acid (H$_2$SO$_4$, Synth), and ammonium molybdate tetrahydrate ((NH$_4$)$_6$Mo$_7$O$_{24}$.4H$_2$O, Sigma-Aldrich) were used as received. All solutions were prepared with ultrapure water obtained with a Millipore Milli-Q system (resistivity > 18 MΩ cm, at 25 °C).

### 2.2. Material Synthesis

#### 2.2.1. Enhancing the NaNbO$_3$ microcubes with WO$_3$ nanoparticle decoration

The NaNbO$_3$ microcubes were synthesized using the methodology outlined by Antonin *et al.* [16]. The synthesized NaNbO$_3$ microcubes underwent a surface modification process involving the deposition of WO$_3$ nanoparticles. It was achieved through an adapted chemical reduction method utilizing ascorbic acid, as detailed in reference [20]. 300 mg of NaNbO$_3$ microcubes were introduced into a 10 mL solution containing 57 mg of polyvinylpyrrolidone and subjected to magnetic stirring for 20 minutes. Subsequently, 10 mL of a solution comprising 568 mg of ascorbic acid was carefully added to the previously prepared mixture while maintaining continuous magnetic stirring.

In a parallel procedure, another solution, consisting of 113 mg of $WCl_4$, was prepared and dissolved in 10 mL of water. This solution was then gradually added dropwise to the solution containing the $NaNbO_3$ microcubes. The resulting mixture was left under magnetic stirring for 24 h. The metal-to-reducing agent-to-directing stabilizer ratio employed in this method was 1:20:2. After the reaction, the final solution underwent centrifugation at 9000 rpm for 10 min and underwent three cycles of washing with water and ethanol. Finally, the microcubes decorated with $WO_3$ nanoparticles were dried in an oven at 100 ºC for 24 h.

*2.2.2. Synthesis of $NaNbO_3$@$WO_3$/C-based electrocatalysts*

Using the impregnation method, NaNbO3 microcubes decorated with WO3 nanoparticles will be supported on carbon[21]. In this procedure, 240 mg of PL6 carbon will be weighed, dispersed in 50 mL of distilled water, and subjected to ultrasonic treatment (UltraSonic cleaner SW2000FI, Sanders) for 20 min. Following this step, 60 mg of $NaNbO_3$@$WO_3$ will be dispersed in 30 mL of water and placed in an ultrasonic bath for 20 min. Subsequently, the dispersion containing the decorated microcubes will be slowly added drop by drop into the dispersion containing the carbon support with continuous magnetic stirring, ensuring that the $NaNbO_3$ microcubes and $WO_3$ nanoparticles are weighed correspondingly to prepare 1%, 3%, 5%, and 10% (wt%) electrocatalysts, maintaining the stirring for 24 h. After this period, the prepared electrocatalysts will be dried at 100 °C in an oven to remove residual water.

*2.3. Material characterization*

The morphology of $WO_3$ was examined through Scanning Electron Microscopy (S.E.M.) utilizing a field emission Scanning Electron Microscope, the FESEM JEOL JSM-6701 F. Semi-quantitative assessments of the mass of the synthesized electrocatalysts, specifically those based on NaNbO$_3$@WO$_3$/C, were acquired from multiple Energy Dispersive Spectroscopy (E.D.S.) mappings conducted on a JEOL JSM SEM-6010LA Scanning Electron Microscope.

X-ray Diffraction (XRD) analyses were conducted utilizing a D8 Focus diffractometer from Bruker A.X.S., equipped with a Cu Kα X-ray source (λ = 1.54 Å). The measurements were performed in continuous scan mode, with a scanning rate of 2 ° per minute, covering a 2$\theta$ range from 10 ° to 80 °.

Contact angle measurements were carried out to assess the hydrophilic/ hydrophobicity properties of the materials under investigation. 4 mg of each electrocatalyst was dispersed in 2 mL of distilled water and subjected to one minute of sonication using an ultrasonic probe system. Subsequently, a 40 µL portion of the dispersion was applied onto a glass plate and allowed to air-dry, forming a uniform thin film. Then, 20 mL of distilled water was added to the dried film to determine contact angles. These measurements were performed in triplicate using Windrop software.

X-ray photoelectron spectroscopy (XPS) examinations were conducted using a Scienta Omicron ESCA + spectrometer from Germany, employing monochromatic Al Kα radiation (1486.7 eV). Shirley's method was applied to subtract the inelastic background from the C 1s high-resolution core-level spectra. The spectra fitting was performed without imposing constraints, employing multiple Voigt profiles through CasaXPS software. The full width at half maximum (F.W.H.M.) ranged between 1.0 and 2.9 eV, and the accuracy of the peak positions was maintained within ± 0.1 eV.

*2.4. Electrochemical measurements*

An Autolab P.G.S.T.A.T. 302N potentiostat was employed for precise potential control. The investigations about the ORR were performed utilizing the Rotating Ring-Disk Electrode (RRDE) technique. For this purpose, an electrode configuration comprising a vitreous carbon disk

(effective surface area of 0.2475 cm$^2$) and a gold ring (effective surface area of 0.1866 cm$^2$), acquired from Pine Instruments and characterized by a collection factor (N) of 0.28, was utilized.

The counter electrode was constituted of Pt, while a Hg|HgO, immersed in a 5.0 M NaOH solution, served as the reference electrode. Rotational speeds were precisely regulated using a CTV101 speed control unit.

The vitreous carbon disk of the commercially available Rotating Disk Electrode served as the substrate for the fabrication of electrocatalysts based on NaNbO$_3$@WO$_3$/C. The synthesis of these electrocatalysts adhered to the methodology proposed by Paulus *et al*. [22]. Before the electrochemical assessments, 100 mL of supporting electrolyte (comprising 1.0 M NaOH) was systematically saturated with molecular oxygen (O$_2$) for 30 minutes. This flow of O$_2$ was consistently maintained throughout all electrochemical measurements, executed at a constant scan rate of 5 mV s$^{-1}$. The polarization curves of all the prepared materials were compared with those obtained with PL6 carbon, the reference material for the 2-electron ORR, and with the platinum curve (E-TEK), the reference material for the 4-electron ORR

### 2.4.1. H$_2$O$_2$ electrogeneration

The RRDE technique was employed to evaluate the electrocatalysts based on their ORR performance and identify the one exhibiting the highest ring current, indicating increased H$_2$O$_2$ electrogeneration. Subsequently, approximately 6 grams of this electrocatalyst were prepared to prepare GDEs. The aim was to quantify the H$_2$O$_2$ electrogeneration facilitated by these electrodes, directly comparing it to the electrogeneration achieved with a PL6 carbon GDE devoid of the addition of metallic nanoparticles.

This synthesis, representing a mass twelve times greater than that used in preparing the electrocatalysts studied using the rotating ring-disk electrode technique (6 grams prepared instead of 0.5 grams), followed a different proportion from that described in section 3.1. It was due to the substantial amount of citric acid and ethylene glycol, resulting in a larger volume of polymer resin, leading to a problematic catalytic mass to macerate. The steps of the polymeric precursor method were followed; however, 10 grams of citric acid (0.05 mol) and 30 mL of ethylene glycol (0.50 mol) were used to prepare the GDEs.

For comparison, GDEs of untreated Printex L6 carbon were prepared using 6 grams of untreated carbon. After measuring the mass of the electrocatalyst, it was transferred with approximately 100 mL of distilled water, and agitation was maintained at 400 rpm for 30 minutes. Subsequently, polytetrafluoroethylene (P.T.F.E., 60% dispersion), acting as a binding agent, was added to the mixture to represent 20% of the final material mass. The agitation was continued for an additional hour. Then, vacuum filtration of this material was performed, and the resulting mass was dried at 110 ºC for 120 minutes.

*2.4.2. Gas diffusion electrode preparation*

Around 6 grams of the electrocatalyst, distinguished for its superior ORR performance as identified through the RRDE technique, were meticulously processed. Furthermore, a GDE composed of unmodified Printex L6 will be prepared to serve as a comparative benchmark. 6 grams of the electrocatalyst will be mixed with 100 mL of ultrapure water and stirred at 400 rpm for 30 minutes. Subsequently, polytetrafluoroethylene (P.T.F.E., 60% dispersion), serving as a binding agent, will be added to the mixture so that P.T.F.E. constitutes 20% of the final material mass, with stirring continued for an additional hour. After this period, the material will undergo vacuum filtration, and the resulting mass will be dried at 110 °C for 120 minutes.

In the assembly process of the GDE, a cylindrical mold with a 2.5 cm diameter is used. The subsequent steps are followed:

(1) firstly, a stainless steel plate with multiple perforations is placed in the mold;

(2) 2 grams of the prepared electrocatalyst are put onto the plate;

(3) a second plate is positioned above the catalytic mass.

The mass is compressed under a force of 2.5 tons for two hours, with the mold temperature maintained at 290 °C.

*2.4.3. Quantification of remaining $H_2O_2$ by subjecting the GDE to electrolysis*

An electrochemical cell was utilized with an electrolyte of $H_2SO_4$ (0.1 M) and $Na_2SO_4$ (0.1 M), pH 3. The electrolyte was saturated with $O_2$ at a pressure of 0.2 bar for 30 min before use, and this flow was maintained for 120 min. To quantify the $H_2O_2$ electrogeneration from the different gas diffusion electrode, extensive electrolysis was performed using -0.7, -1.1, -1.5, -1.9, and -2.3 V. Pt and Ag|AgCl were utilized as the counter and reference electrodes, respectively.

During electrolysis, 500 µL of sample was collected from the cell and added to 4 mL of ammonium molybdate solution (2.4 x $10^{-3}$ M in 0.5 M $H_2SO_4$) to form a yellow peroxymolybdate complex. These solutions were analyzed using a Varian Cary 50 spectrophotometer ($\lambda$ = 350 nm) to determine the concentration of electrogenerated residual $H_2O_2$.

From the electrical current (i, in A) and cell potential ($E_{cell}$, in V) measured using the most effective $NaNbO_3$@$WO_3$/C-based material for ORR GDE and a Printex L6 GDE, a comparative assessment of energy consumption (E.C., in kW h $kg^{-1}$) was conducted following equation (2) [23].

$$EC = \frac{i\, E_{cel}\, t}{m} \quad (2)$$

where t is the electrolysis time (in h), and m is the mass of $H_2O_2$ produced (in kg).

Hence, the calculation of current efficiency (C.E.) for $H_2O_2$ generation, signifying the ratio of electrical energy consumed by the electrode reaction to the overall electrical energy passing through the circuit, was determined through the utilization of equation (3) [24]:

$$CE(\%) = \frac{nFC_{H_2O_2}V}{It} \times 100\% \qquad (3)$$

where $n$ is the number of electrons transferred for oxygen reduction to $H_2O_2$, $F$ is the Faraday constant (96,486 C mol$^{-1}$), $C_{H_2O_2}$ is the concentration of $H_2O_2$ (mol L$^{-1}$), $V$ is the electrolyte volume (L), $I$ is the applied current intensity (A), and $t$ is the reaction time (s).

## 3. Results and discussion

*3.1. The synthesis of sodium niobate (NaNbO₃) microcubes decorated with WO₃ nanoparticles*

The SEM-FEG analysis was conducted to investigate the morphological properties of NaNbO₃@WO₃ particles, and the acquired images are displayed in **Fig. 1A**, confirming the formation of NaNbO₃ microcubes and the presence of reduced WO₃ on the microcube surfaces and in isolation. Some microcubes were entirely adorned with WO₃ particles, while others displayed random distribution. It is possible to observe that the WO₃ particles are of nanometric dimensions, while the microcubes display micrometric dimensions.

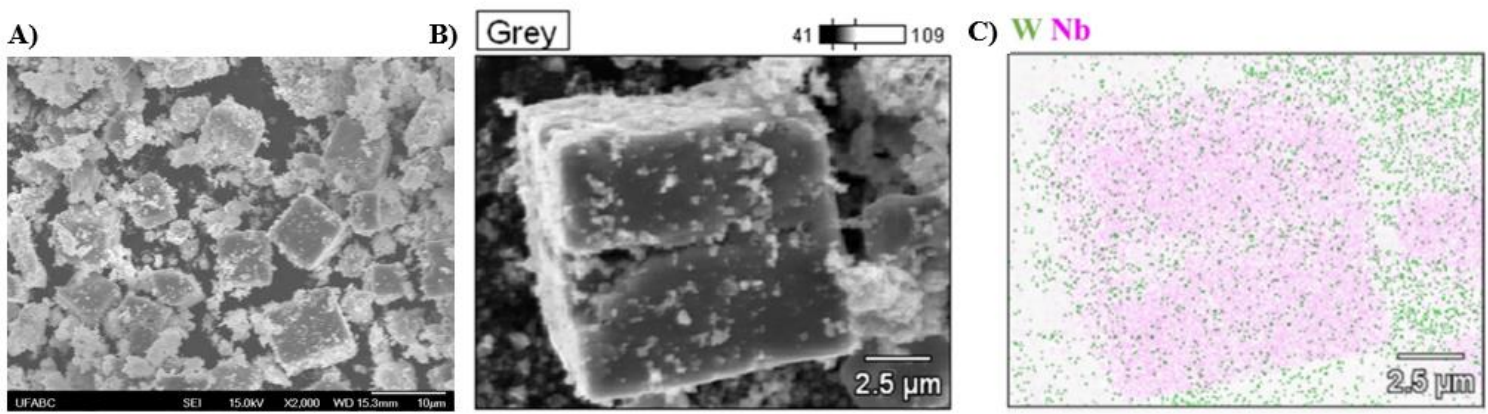

**Figure 1.** SEM-FEG image of (A) NaNbO$_3$ microcubes decorated with WO$_3$ nanoparticles, (B) an image of a single NaNbO$_3$ microcube decorated with WO$_3$ nanoparticles, and (C) E.D.S. mapping of the individual NaNbO$_3$ microcube depicted in Fig. 1B.

In **Fig. 1B**, an image of a single decorated microcube is presented, revealing that the WO$_3$ nanoparticles cover the NaNbO$_3$ microcubes in a non-aggregated, random manner. This arrangement exposes catalytic sites on the NaNbO$_3$ microcube and the WO$_3$ nanoparticles. In **Fig. 1C**, an elemental mapping of W (green points) and Nb (pink points) is displayed using E.D.S. from S.E.M. This provides evidence that the decoration of the NaNbO$_3$ microcubes was successful, as indicated by the earlier images. Furthermore, it confirms the presence of tiny clusters of WO$_3$ nanoparticles around the NaNbO$_3$ microcubes.

Furthermore, the E.D.S. line scan mapping was conducted from the outer cluster towards the center of the microcubes, as depicted in **Fig. 2**. According to **Fig. 2B**, elements Na, O, Nb, and W were identified, with the concentration of W decreasing as the E.D.S. scan progresses towards the microcube's center, while Nb and Na concentrations increase. The concentration of O remains relatively stable along the measurement arrow. This outcome provides a consistent indication that the cluster on the surface of the NaNbO$_3$ microcube is composed of WO$_3$ nanoparticles.

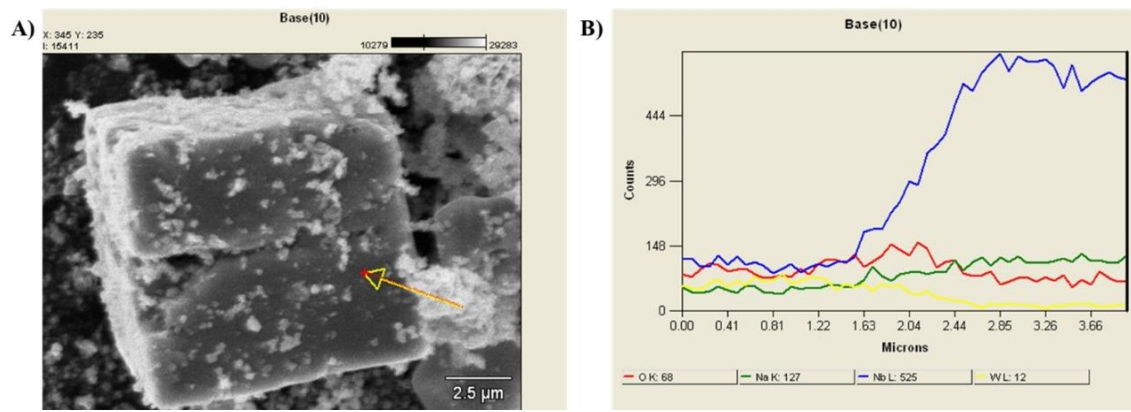

**Figure 2.** An E.D.S. line scan mapping of the NaNbO$_3$@WO$_3$ material was obtained via SEM-FEG. (A) Illustrates the line scan measurement arrow, (B) depicts the micrometer scale along the arrow in Fig. 3A, and the element presence counts.

**Fig. 3** presents another image of the NaNbO$_3$ microcube decorated with WO$_3$ (NaNbO$_3$@WO$_3$), further substantiating the evidence discussed thus far. A few clusters of WO$_3$ nanoparticles, measuring approximately 1 micron in size, can also be observed. E.D.S. mapping with atomic proportion calculation and E.D.S. spectrum analysis were performed from this image. The mapping reveals the presence of Na and Nb, with the microcube's shape, while the presence of W is more pronounced at the edges of the NaNbO$_3$ microcube. Additionally, oxygen is detected in both structures.

The E.D.S. spectrum strengthens the evidence that the elements above are proper. However, the atomic proportion of Na, Nb, O, and W is localized, representing a relatively small portion of the sample. The atomic proportions of Nb and W, at 30:26, are close to the expected 1:1 synthesis ratio, indicating that the composition aligns with the synthesis expectations. The atomic proportions of O and Na are 36 and 7, respectively, with the high presence of O being anticipated since this element is integral to both structures under investigation.

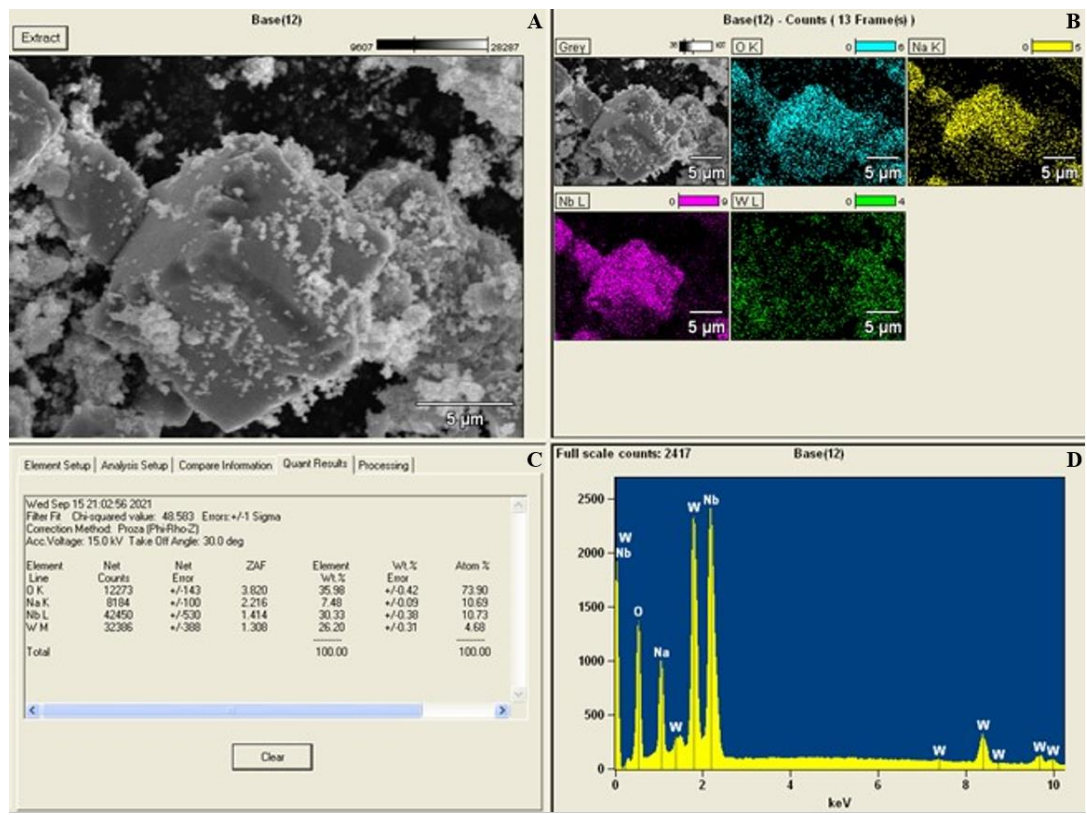

**Figure 3.** (A) Image of NaNbO$_3$ microcubes decorated with WO$_3$ obtained by SEM-FEG. (B) Color-coded E.D.S. mapping of the NaNbO$_3$@WO$_3$ material corresponding to the image in Fig. 4A. (C) Table of element concentrations, and (D) E.D.S. spectrum.

The catalysts' crystalline structures were analyzed through XRD. **Fig. 4** displays the XRD patterns of NaNbO$_3$/C and NaNbO$_3$@WO$_3$/C. The broad peaks observed at $2\theta$ = 12.1 ° and 25.1 ° are consistent across all catalyst samples and are attributed to the carbon. The XRD pattern of NaNbO$_3$/C reveals a cubic crystal structure with the Pm-3m space group, following the corresponding entry in the Inorganic Crystal Structure Database (I.C.S.D.) numbered 192408, with no detectable undesirable signal phases. In the patterns of NaNbO$_3$@WO$_3$/C samples, the NaNbO$_3$ signals were preserved; three peaks related to WO$_3$ oxides were observed in 10% NaNbO$_3$@WO$_3$/C sample, the peaks related to the orthorhombic crystal structure appear at $2\theta$ = 16.5 ° and 25.7 °, additionally in $2\theta$ = 28.0 ° was observed a peak related to hexagonal structure. The XRD patterns 1-5 % NaNbO$_3$@WO$_3$/C did not exhibit any observed diffraction peaks for crystalline WO$_3$, indicating that WO$_3$ species were highly dispersed on the surface of NaNbO$_3$/C

[25]. Some work reports that phase junction has been beneficial in improving the efficiency of catalytic applications [26–28].

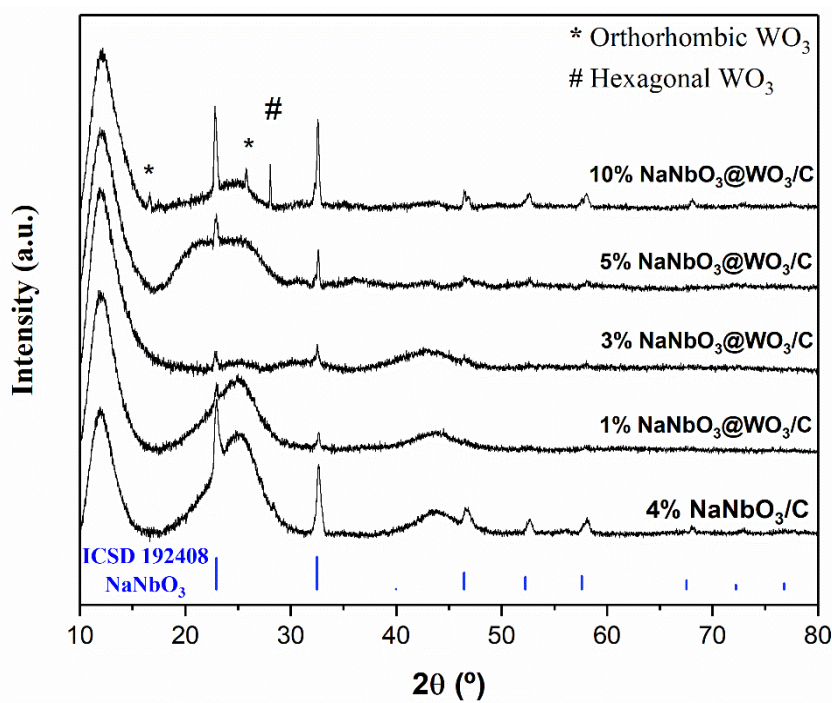

**Figure 4.** (a) XRD patterns of 4% NaNbO$_3$/C and NaNbO$_3$@WO$_3$/C-based materials.

To evaluate the hydrophilicity of the synthesized materials, we determined the average contact angle values of NaNbO$_3$@WO$_3$/C, 4% NaNbO$_3$/C, and unmodified Printex L6 carbon, which are summarized in **Table 1**. The results show that compared to unmodified Printex L6 carbon, both NaNbO$_3$@WO$_3$/C-based materials and 4% NaNbO$_3$/C electrocatalysts exhibit higher hydrophilicity, as demonstrated by their lower contact angle (approximately 20°), resulting in better wettability. This phenomenon can be attributed to oxy acid functional groups on the modified Printex L6 carbon [8,29,30], but in an amount lower than with the materials containing Niobate and tungsten oxide, a finding further confirmed by subsequent XPS analysis.

**Table 1.** Average contact angles of NaNbO$_3$@WO$_3$/C-based materials, 4% NaNbO$_3$/C, and Printex L6 without modification, along with the corresponding standard errors obtained for each material.

| Electrocatalyst | Contact angle (°) |
|---|---|
| 1% NaNbO$_3$@WO$_3$/C | 21 ± 2 |
| 3% NaNbO$_3$@WO$_3$/C | 19 ± 2 |
| 5% NaNbO$_3$@WO$_3$/C | 19 ± 1 |
| 10% NaNbO$_3$@WO$_3$/C | 25 ± 1 |
| 4% NaNbO$_3$/C | 20 ± 2 |
| Printex L6 | 45 ± 3 |

Hydrophilic materials have achieved notable outcomes in prior studies, particularly in H$_2$O$_2$ electrogeneration [8,31,32]. Among various investigations and experimental tests, an optimal oxygenated species content in the range of 15-17%, representing a 5-10% increase in oxygenated functional groups compared to pristine carbon, has been identified as conducive to enhancing the reaction. While it cannot be definitively asserted that surface hydrophilicity is the sole determinant of catalytic performance, it is considered an ancillary parameter in our comprehensive analysis. Based on the literature, it is conceivable that additional factors contribute to the notable reactivity observed in hydrophobic materials. Such elements may encompass variations in material structure, such as carbon fiber [33] compared to carbon powder (this study), distinct functionalization approaches, such as N-doping [34] versus metal oxide modification (this study), and so far. Conversely, Zhang *et al*. [35] revealed that a Janus electrode, characterized by its asymmetric wettability structure featuring a hydrophobic gas storage layer and a hydrophilic catalyst layer, not only ensures an adequate oxygen supply but also leads to increased rates of H$_2$O$_2$ generation and enhanced oxygen utilization efficiency.

XPS analyses were carried out for the PL6 and 5% NaNbO$_3$@WO$_3$/C samples to investigate the presence of functional groups. The high-resolution C1s spectra presented in **Fig. 5** were deconvolved into four components related to different carbon chemical environments.

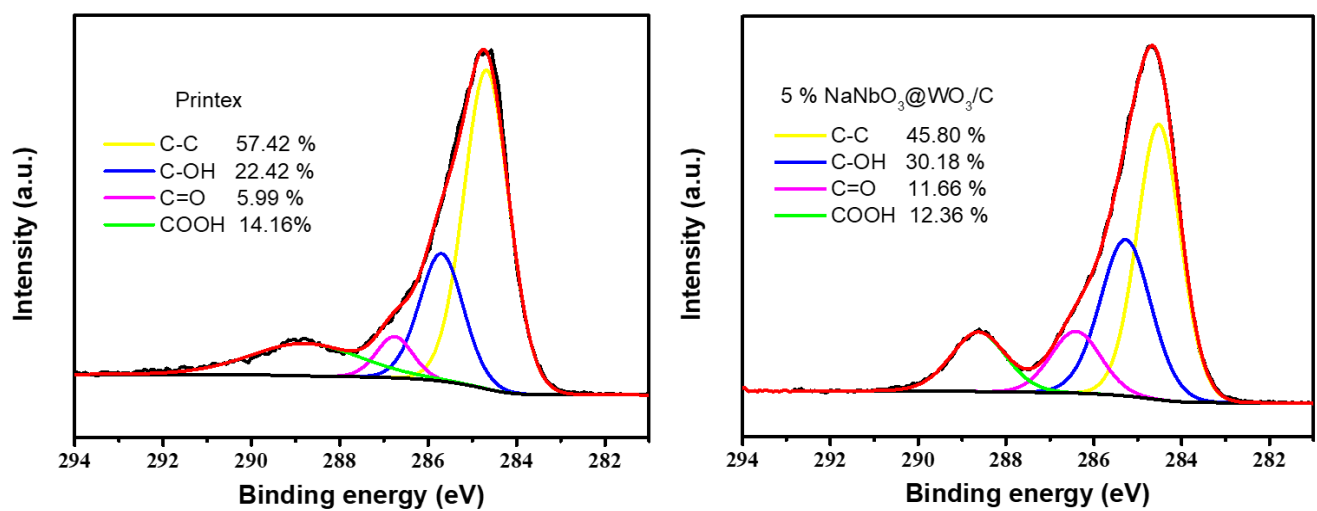

**Figure 5.** Deconvoluted C1s XPS spectra of Printex and 5% NaNbO$_3$@WO$_3$/C.

The peak at approximately 284 eV was assigned to groups C-C, while the peaks at approximately 285, 286, and 289 eV were assigned to the C-OH, C=O, and -C.O.O.H. groups, respectively [36,37]. It can be observed that the 5 % NaNbO$_3$@WO$_3$/C sample presented 54.2 at. % of functional groups containing oxygen while the PL6 sample presented 42.6 at. %. Oxygenated species on the surface of catalysts can increase their hydrophilic character and facilitate mass transfer between dissolved oxygen and the catalyst surface. In this way, the electrogeneration of H$_2$O$_2$ is favored in catalysts with a surface rich in oxygen-containing functional groups [6,38].

*3.2. Electrochemical measurements*

Following the physical characterization of the materials, an evaluation of their electrocatalytic performance for the ORR was undertaken. The ORR polarization curves are depicted in **Fig. 6**. This analysis included 4% $NaNbO_3$/C, $NaNbO_3@WO_3$/C-based materials, and unmodified carbon (Printex L6), serving as a conventional reference material for the 2-electron mechanism. Furthermore, the outcomes of 20% Pt/C, established as a reference material for the 4-electron mechanism, have also been incorporated for comparative purposes.

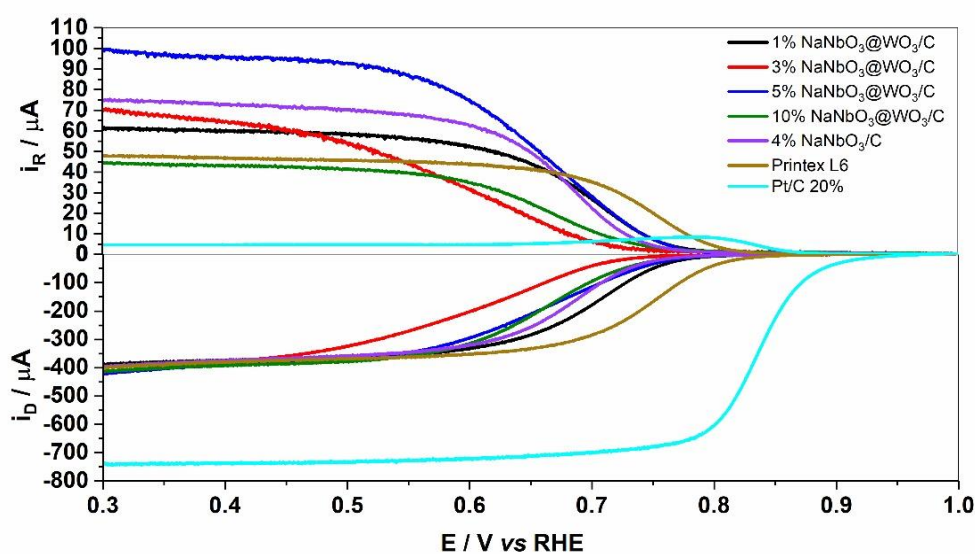

**Figure 6.** A comparative analysis of stable-state polarization curves resulting from the ORR was conducted in an $O_2$-saturated 1.0 mol $L^{-1}$ NaOH electrolyte at 25 ºC, employing a rotation rate of 1600 rpm and the RRDE system. The working electrodes were prepared using a thin Printex L6-based materials layer under different conditions: unmodified (——), with 1% $NaNbO_3@WO_3$/C (——), with 3% $NaNbO_3@WO_3$/C (——), with 5% $NaNbO_3@WO_3$/C (——), with 10% $NaNbO_3@WO_3$/C (——), with 4% $NaNbO_3$/C (——), and with 20% Pt/C (——). The polarization curves were generated with a scan rate of 5 mV $s^{-1}$, illustrating (A) ring currents (polarized at 0.3 V) and (B) disk currents.

The positive current value (**Fig. 6A**) is directly associated with the production of $H_2O_2$ at the disk electrode and its subsequent oxidation at the ring electrode. The ring current exhibits the following order from highest to lowest: 5% $NaNbO_3@WO_3$/C > 4% $NaNbO_3$/C > 3% $NaNbO_3@WO_3$/C > 1% $NaNbO_3@WO_3$/C, with Printex L6, 10% $NaNbO_3@WO_3$/C, and 20% Pt/C following in sequence. Therefore, among the tested electrocatalysts, the one that proved to be the most promising for $H_2O_2$ electrogeneration was the 5% $NaNbO_3@WO_3$/C. The RRDE on

Pt/C E-TEK exhibits a high current at the disk and a low current at the ring, indicating a preference for the 4-electron mechanism. On the other hand, the RRDE on Printex L6 carbon shows a lower current at the disk than when using Pt/C E-TEK and a higher current at the ring, indicating a greater preference for the 2-electron mechanism. The polarization curves of all NaNbO$_3$@WO$_3$/C-based materials exhibit similar profiles, and the disk currents (**Fig. 6B**) closely resemble those achieved with Printex L6 carbon, suggesting the potential occurrence of a 2-electron ORR mechanism.

By plotting $i^{-1}$ x $\omega^{\frac{1}{2}}$ (where $i$ is the measured current and $\omega$ is the angular speed of rotation of the electrode in rpm) as a first-degree function, a linear relationship is established with y = ax + b. The slope of this line is represented as $\frac{1}{B}$, can be used to determine the number of electrons involved in the electrochemical process. This technique is known as the Koutecky-Levich plot and is particularly valuable for quantifying the number of electrons at a specific potential. When evaluating electrocatalysts, the calculated slope is compared to carbon and platinum. If the value closely resembles that of Printex L6 carbon, the substance likely favors the 2-electron mechanism of ORR, resulting in H$_2$O$_2$ production. Conversely, if the electrocatalyst's slope is similar to platinum's, it indicates activity in generating H$_2$O via the 4-electron mechanism. The Koutecky-Levich plot for all the electrocatalysts developed in this study is presented in **Fig. S1**.

Parallel lines for all materials were determined using the Koutecky-Levich equation. The angular coefficients for the various materials, such as 1% NaNbO$_3$@WO$_3$/C (349 (rpm)$^{-1/2}$ (mA)$^{-1}$), 3% NaNbO$_3$@WO$_3$/C (342 (rpm)$^{-1/2}$ (mA)$^{-1}$), 5% NaNbO$_3$@WO$_3$/C (332 (rpm)$^{-1/2}$ (mA)$^{-1}$, 10% NaNbO$_3$@WO$_3$/C (366 (rpm)$^{-1/2}$ (mA)$^{-1}$, and 4% NaNbO$_3$/C (rpm)$^{-1/2}$ (mA)$^{-1}$, closely resemble the value obtained for Printex L6 carbon (rpm)$^{-1/2}$ (mA)$^{-1}$, suggesting a 2-electron transfer mechanism. In contrast, the angular coefficient of the 20% Pt/C material is significantly higher at 1225 (rpm)$^{-1/2}$ (mA)$^{-1}$, indicating a 4-electron ORR mechanism. Printex L6 carbon catalysts primarily yield H$_2$O$_2$ during ORR, with a 2-electron transfer per oxygen molecule. The R.D.D.E. experiments in this study suggest that NaNbO$_3$@WO$_3$/C-based catalysts likely exhibit a similar catalytic profile.

The kinetic current density for ORR was calculated from the Koutecky–Levich equation [39,40]. Among all electrocatalysts, the 5% NaNbO$_3$@WO$_3$/C displayed the most favorable kinetic current density, as depicted in **Fig 7**.

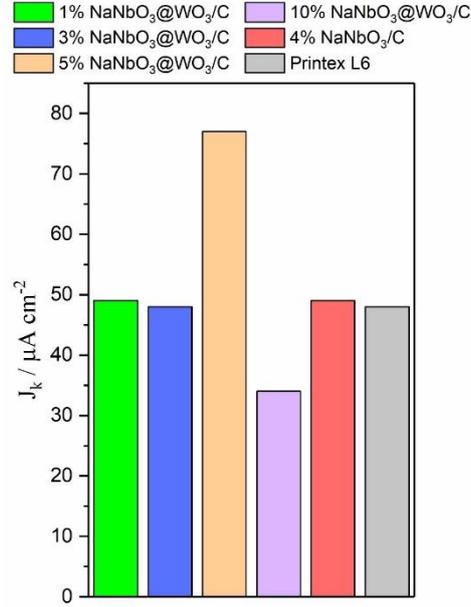

**Figure 7.** Kinetic current density from ORR from Koutecky–Levich plots.

The percentage of H$_2$O$_2$ generated (p(H$_2$O$_2$)) and the number of electrons transferred by each O$_2$ molecule ($n_t$) were calculated following the methodologies outlined in the studies by Demarconnay *et al*. [41] and Jakobs *et al*. [42], using equations 2-4. The results can be found in **Table 2**.

$$p(H_2O) = \frac{N\left(-\frac{j_D}{j_{R,1}-j_{R,1}^0}\right)-1}{N\left(-\frac{j_D}{j_{R,1}-j_{R,1}^0}\right)+1} \tag{2}$$

$$p(H_2O_2) = 100 - p(H_2O) \tag{3}$$

$$n_t = 2[p(H_2O)+1] \tag{4}$$

where N is the collection efficiency, $j_D$ is the disk current density, $j_{R,1}$ is the limiting ring current density for disk potentials ($E_{D.}$) lower than 0.9 V versus R.H.E. and $j_{R,1}^0$ is the limiting ring current density for disk potentials ($E_{D.}$) higher than 0.9 V versus R.H.E.

**Table 2.** Percentage of electrogenerated $H_2O_2$ and $H_2O$, and the electron transfer number for electrocatalysts.

| Electrocatalyst | Number of electrons | % $H_2O$ | % $H_2O_2$ |
|---|---|---|---|
| 1% NaNbO$_3$@WO$_3$/C | 2.5 | 28 | 72 |
| 3% NaNbO$_3$@WO$_3$/C | 2.5 | 26 | 74 |
| 5% NaNbO$_3$@WO$_3$/C | 2.1 | 6 | 94 |
| 10% NaNbO$_3$@WO$_3$/C | 2.9 | 44 | 56 |
| 4% NaNbO$_3$/C | 2.3 | 18 | 82 |
| Printex L6 | 2.7 | 38 | 62 |
| 20% Pt/C | 3.9 | 94 | 6 |

The data presented in **Table 2** reveals that the 5% NaNbO$_3$@WO$_3$/C electrocatalyst exhibited the most pronounced activity in $H_2O_2$ generation. Furthermore, this electrocatalyst displayed the highest ring current and accomplished 2.1 transferred electrons in ORR per $O_2$ molecule, resulting in an electrogeneration rate of approximately 94% for $H_2O_2$. In contrast, Printex L6, without any modifications, yielded 62% $H_2O_2$ and transferred 2.7 electrons. Given its promising performance ORR for $H_2O_2$ electrogeneration, the 5% NaNbO$_3$@WO$_3$/C material is now being further characterized for macroscale applications by utilizing GDEs.

### 3.2.1. Performance of electrocatalytic $H_2O_2$ synthesis

The electrochemical production of $H_2O_2$ on GDE utilizing 5% NaNbO$_3$@WO$_3$/C and Printex L6 carbon without modification is displayed in **Fig. 8A**. **Fig. 8B** illustrates the energy consumption necessary for generating the equivalent $H_2O_2$ quantity through GDEs in correlation with the employed voltage. Additionally, current efficiency for $H_2O_2$ production is depicted in **Fig. 8C**.

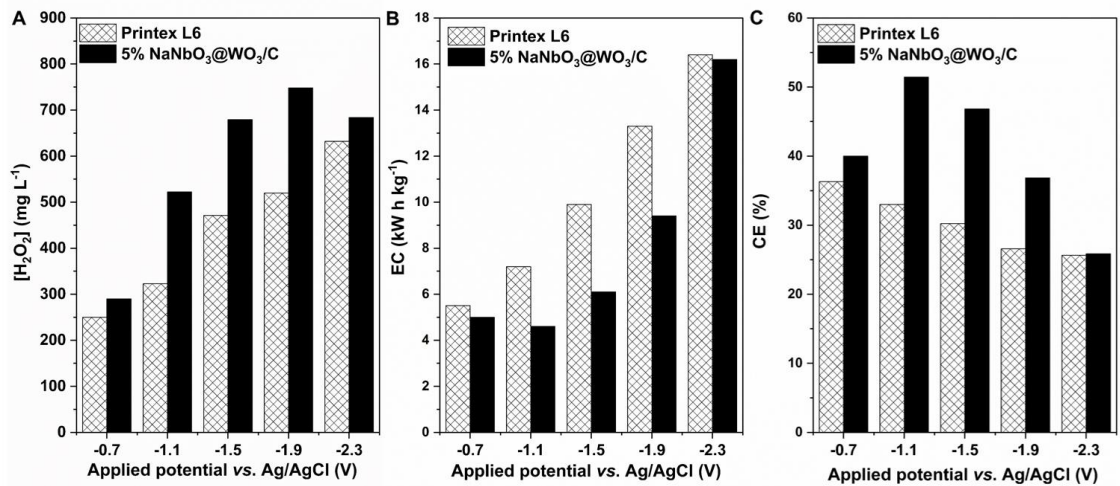

**Figure 8.** (A) Concentration of $H_2O_2$ produced by pure Printex L6 carbon GDE and by Printex L6 carbon modified with 5% $NaNbO_3$@$WO_3$ GDE (B) Energy consumption about the applied current during the electrolysis. (C) Current efficiency for the generation of $H_2O_2$. The measurements were taken after a 2-hour electrolysis period with a constant applied potential, employing a 0.1 M $Na_2SO_4$ + 0.1 M $H_2SO_4$ electrolyte at pH 3, electrochemical cell volume = 350 mL.

**Fig. 8A** shows that $H_2O_2$ electrogeneration increases with higher applied voltages for both materials, except upon utilization of 5% $NaNbO_3$@$WO_3$/C GDE at a potential of -2.3 V. The substantial enhancement in $H_2O_2$ production with the modified catalyst becomes distinctly noticeable in the potential range from -1.1 V to -1.9 V. At -1.1 V, -1.5 V, and -1.9 V, the Printex L6 GDE generated 323, 471 and 520 mg $L^{-1}$ of $H_2O_2$, respectively. At the same time, the 5% $NaNbO_3$@$WO_3$/C GDE produced almost twice as much $H_2O_2$ (522, 679, and 748 mg $L^{-1}$, respectively) under identical conditions, highlighting the enhanced $H_2O_2$ electrogeneration of Printex L6 carbon modified with 5% $NaNbO_3$@$WO_3$/C. Therefore, it can be observed that the disparity in $H_2O_2$ electrogeneration between the 5% $NaNbO_3$@$WO_3$/C GDE and the Printex L6 carbon GDE is directly proportional. When applying a potential of -1.1 V, the difference amounted to 199 mg $L^{-1}$; at -1.5 V, it reached 208 mg $L^{-1}$. At -1.9 V, the difference was 228 mg $L^{-1}$. Hence, despite the increased energy demand, a nearly proportional increase in electrogeneration was observed, indicating the need for additional energy consumption and current efficiency calculations to establish the optimal parameters for utilizing the 5% $NaNbO_3$@$WO_3$ GDE

Contrasting the calculated energy consumption (**Fig. 8B**), it can be observed that the 5% $NaNbO_3$@$WO_3$/C GDE is more efficient from an economic perspective when applying -1.1, -1.5, and -1.9 V for $H_2O_2$ electrogeneration requiring notably less energy. On the other hand, the Printex L6 carbon GDE yields a lower $H_2O_2$ concentration, demanding more energy consumption. However, for the 5% $NaNbO_3$@$WO_3$/C GDE, when potentials of -0.7 V and -2.3 V are applied, the energy demands are pretty close to those of using the Printex L6 GDE under the same conditions. The energy consumption when potentials of -1.1 V (4.6 kW h $kg^{-1}$ for the 5% $NaNbO_3$@$WO_3$/C GDE and 7.2 kW h $kg^{-1}$ for the Printex L6 GDE), -1.5 V (6.1 kW h $kg^{-1}$ for the 5% $NaNbO_3$@$WO_3$/C GDE and 9.9 kW h $kg^{-1}$ for the Printex L6 GDE), and -1.9 V (9.4 kW h $kg^{-1}$ for the 5% $NaNbO_3$@$WO_3$/C GDE and 13.3 kW h $kg^{-1}$ for the Printex L6 GDE), the difference in consumption between the two electrodes corresponds to 2.6, 3.8, and 3.9 kW h $kg^{-1}$, respectively. Therefore, -1.5 and -1.9 V potentials appear more economically viable for utilizing the 5% $NaNbO_3$@$WO_3$/C GDE, as they generate more $H_2O_2$ while consuming less energy than the Printex L6 GDE

Also, the current efficiency (**Fig. 8C**) when employing the 5% $NaNbO_3$@$WO_3$/C GDE and applying a potential of -1.5 V reached approximately 47%, compared to nearly 30% when utilizing the Printex L6 GDE under identical conditions. When -1.9 V was applied, C.E. was 37% for the 5% $NaNbO_3$@$WO_3$/C GDE and 26% for the Printex L6 GDE Consequently; it can be deduced that the best potential for $H_2O_2$ electrogeneration with 5% $NaNbO_3$@$WO_3$/C electrocatalyst is -1.5 V, resulting in an electrogeneration of 679 mg $L^{-1}$ with an energy consumption of approximately 6 kW h $kg^{-1}$ and current efficiency of 47%. For the reasons pointed out above, the 5% $NaNbO_3$@$WO_3$/C electrocatalyst has higher hydrogen peroxide electrogeneration, lower energy consumption, and higher C.E. than Printex L6 carbon.

The remarkable enhancement in catalytic activity observed during the $H_2O_2$ electrogeneration process with the utilization of the 5% $NaNbO_3$@$WO_3$/C electrocatalyst is inherently linked to its distinctive property of efficiently facilitating electron transfer throughout key reaction steps [16]. This pivotal characteristic, the proficient transfer of electrons across the catalytic surface, underscores the catalytic prowess of $NaNbO_3$, particularly evident in its efficacy in the 2-electron ORR Significantly, many research endeavors focused on $WO_3$ deposition on diverse supports have substantiated the observation that this catalyst manifests surface acidity attributed to the existence of both Lewis and Brønsted acidic sites associated with $W^{6+}$ species [43]. Notably, the acidic surface displays an increased attraction to species with $O.H.^-$, thereby enhancing oxygen concentrations and promoting the production of $H_2O_2$. In conclusion, the synergy between the unique properties of $NaNbO_3$ and $WO_3$ within the 5% $NaNbO_3@WO_3/C$ electrocatalyst underscores its potential as a high-performance material for catalyzing ORR, offering enhanced catalytic activity, making it a promising candidate for various electrochemical applications.

A comprehensive comparison with some studies verified that electrocatalysts derived from tungsten oxide, nickel, and cerium dioxide demonstrate significant promise for $H_2O_2$ electrogeneration [4]. The GDE with 1% $WO_3/C$ produced 585 mg $L^{-1}$ of $H_2O_2$ at a potential of -1.3 V [19]. At the same time, the GDE with $WO_{2.72}/Vn$ achieved notable performance in $H_2O_2$ generation at its optimal potential of -0.7 V vs Ag/AgCl, producing approximately 480 mg $L^{-1}$ of $H_2O_2$ with an energy consumption value of 13.8 kWh $kg^{-1}$ [32]. Thus, using a 5% $NaNbO_3@WO_3/C$ electrocatalyst resulted in a higher $H_2O_2$ generation rate (679 mg $L^{-1}$) than other reported literature works.

## 4. Conclusions

Electrocatalysts prepared using 1%, 3%, and 5% loadings of $NaNbO_3$@$WO_3$/C demonstrated superior effectiveness and selectivity in $H_2O_2$ electrogeneration compared to unmodified Printex L6. Among these, the 5% $NaNbO_3$@$WO_3$/C catalyst proved the most efficient, achieving a higher ring current and catalyzing the conversion of 94% of $O_2$ to $H_2O_2$. Furthermore, it facilitated a 2.1 electron transfer in the ORR On the other hand, Printex L6 had a selectivity of only 62% to $H_2O_2$ and transferred 2.7 electrons in ORR The 5% $NaNbO_3$@$WO_3$/C electrocatalyst has higher hydrogen peroxide electrogeneration, lower energy consumption, and higher C.E. than Printex L6 carbon.

The superior performance of $NaNbO_3$@$WO_3$/C-based electrocatalysts stems from the efficient electron transfer facilitated by $NaNbO_3$ during crucial steps. Concurrently, the deposition of $WO_3$ unveils surface acidity, enhancing oxygen concentrations and elevating hydrophilicity. This effect arises from a heightened concentration of acidic oxygenated species on the surface, as confirmed by contact angle measurements and XPS analysis. These electrocatalysts hold significant promise for *in situ* $H_2O_2$ generation and future applications in electrochemical advanced oxidation processes for degrading organic pollutants.

**Acknowledgments**

The authors would like to thank F.A.P.E.S.P. (2017/21846-6, 2020/14100-0, 2021/05364-7, 2021/14394-7, 2022/10484-4), CAPES (88887.354751/2019-00) and CNPq (303943/2021-1, 310045/2019-3, 429727/2018-6, 402609/2023-9) for the financial support.